\documentclass[pra,aps,10pt,superscriptaddress,twocolumn,floatfix]{revtex4-1}
\usepackage{graphicx,color}
\graphicspath{{figpdf_paper/}}
\usepackage[nice]{nicefrac}							
\usepackage{amsmath,amssymb,bm}
\usepackage{braket}

\newcommand{\ii}{\mathrm{i}} 

\newcommand{\ee}{e}

\usepackage[plainpages=false,pdfpagelabels,colorlinks=true,linkcolor=red,urlcolor=blue,citecolor=blue,pdftitle={},pdfauthor={},pdfdisplaydoctitle=true,pdfduplex=DuplexFlipLongEdge]{hyperref}

\definecolor{darkred}{rgb}{0.90,0.2,0.2}
\definecolor{darkgreen}{rgb}{0,0.60,.2}
\definecolor{darkblue}{rgb}{0.1,0.3,1}
\definecolor{grey}{cmyk}{0,0,0,0.25}
\definecolor{orange}{cmyk}{0,0.6,0.8,0}

\begin{document}

\title{Localization challenges quantum chaos in the finite two-dimensional Anderson model}

\author{Jan \v Suntajs}
\affiliation{Department of Theoretical Physics, J. Stefan Institute, SI-1000 Ljubljana, Slovenia}
\affiliation{Department of Physics, Faculty of Mathematics and Physics, University of Ljubljana, SI-1000 Ljubljana, Slovenia\looseness=-1}
\author{Toma\v z Prosen}
\affiliation{Department of Physics, Faculty of Mathematics and Physics, University of Ljubljana, SI-1000 Ljubljana, Slovenia\looseness=-1}
\author{Lev Vidmar}
\affiliation{Department of Theoretical Physics, J. Stefan Institute, SI-1000 Ljubljana, Slovenia}
\affiliation{Department of Physics, Faculty of Mathematics and Physics, University of Ljubljana, SI-1000 Ljubljana, Slovenia\looseness=-1}


\begin{abstract}
It is believed that the two-dimensional (2D) Anderson model exhibits localization for any nonzero disorder in the thermodynamic limit and it is also well known that the finite-size effects are considerable in the weak disorder limit.
Here we numerically study the quantum-chaos to localization transition in the finite 2D Anderson model using standard indicators used in the modern literature such as the level spacing ratio, spectral form factor, variances of observable matrix elements, participation entropy and the eigenstate entanglement entropy.
We show that many features of these indicators may indicate emergence of robust single-particle quantum chaos at weak disorder.
However, we argue that a careful numerical analysis is consistent with the single-parameter scaling theory and predicts the breakdown of quantum chaos at any nonzero disorder value in the thermodynamic limit. Among the hallmarks of this breakdown are the universal behavior of the spectral form factor at weak disorder, and the universal scaling of various indicators as a function of the parameter $u = \left(W \ln V\right)^{-1}$ where $W$ is the disorder strength and $V$ is the number of lattice sites.
\end{abstract}

\maketitle

\section{Introduction}
Identifying regimes in which quantum dynamics exhibits chaotic behavior is a nontrivial task and is currently subject of active research in different fields of condensed matter~\cite{dalessio_kafri_16, bertini_heidrichmeisner_21}, quantum statistical physics~\cite{dalessio_kafri_16} and quantum information~\cite{PRXQuantum.3.030201}.
One of the central questions is to understand the general circumstances under which quantum chaos~\cite{stockmann2000quantum, haake1991quantum} and eigenstate thermalization~\cite{deutsch_91, srednicki_94, rigol_dunjko_08, dalessio_kafri_16} break down.
In a given microscopic theory, one may distinguish between two types of transitions between quantum chaos and the absence thereof in the thermodynamic limit:
a case in which one of the regimes is limited to a singular point in a parameter space, and a more general case that exhibits a true transition between chaotic and non-chaotic regime.

A paradigmatic example of the breakdown of single-particle quantum chaos in lattice systems is the Anderson model, which at sufficiently large disorder hosts the well-known Anderson localization~\cite{anderson_58}. It was realized rather soon that one-dimensional systems localize for any nonzero disorder amplitude $W$~\cite{Mott1961,R.E.Borland1963}, and in three (and more) dimensions the localization transition should emerge at some nonzero $W$~\cite{anderson_58}.
Interestingly, in 1980's the problem of localization in the Anderson model in $D$ dimensions has been mapped to the problem of suppression of diffusion by dynamical localization in the standard model of quantum chaos, the so-called kicked rotator~\cite{fishman}, with $D$ incommensurate frequencies, which has been specifically studied for $D=2$~\cite{shepelyansky} and $D=3$~\cite{casati}.

The situation, however, turned out to be more challenging in two dimensions (2D).
It is now believed that in 2D, much as in one dimension, localization emerges in the thermodynamic limit at any nonzero disorder~\cite{abrahams_anderson_79}, and hence delocalization is only limited to a singular point at disorder $W=0$. Intriguingly, however, finite 2D systems appear to exhibit a relatively broad regime at weak disorder in which fingerprints of localization appear to be less clear.
As a consequence of this ambiguity, in the period from 1958 (introduction of the Anderson model~\cite{anderson_58}) until around 1979 (introduction of the single-parameter scaling theory~\cite{abrahams_anderson_79}), it was commonly argued that the 2D systems exhibit a localization transition at some nonzero disorder~\cite{edwards_thouless_72, licciardello_economou_75, licciardello_thouless_75, licciardello_thouless_75b, yoshino_okazaki_77, weaire_srivastava_77, prelovvsek1978diffusion, PhysRevLett.40.1596, PhysRevLett.42.1492, stein_krey_80, kramer1981numerical}. 
The origin of the latter confusion can be attributed to the exponential scaling of the localization length with the inverse disorder amplitude~\cite{mackinnon_kramer_81, mackinnon1983scaling, economou_soukoulis_84, schreiber_ottomeier_92, manai_clement_15, chakraborty_gorantla_20}, such that at very weak disorder even macroscopically large samples may not localize~\cite{licciardello_thouless_78, mackinnon1983scaling}.
The validity and the level of applicability of the single-parameter scaling theory then remained a matter of discussion~\cite{Kaveh_1981, pichard_sarma_81a, pichard_sarma_81b, pichard1985power, Schreiber_1985}.

This work combines two viewpoints on this long-standing problem.
In the first, rather historical perspective, we investigate the weak disorder regime of the 2D Anderson model through the lens of of spectral statistics, matrix elements of observables, and the structure of wavefunctions.
We show that the results in finite systems appear to be compatible with the emergence of single-particle quantum chaos~\cite{lydzba_rigol_21, PhysRevB.104.214203} at weak disorder.
However, we also find simple and powerful scaling solutions that are consistent with single-parameter scaling theory~\cite{abrahams_anderson_79}, and demonstrate the flow towards a localized state at any nonzero disorder in the thermodynamic limit.

The second perspective is aligned with concurrent studies of the breakdown of quantum chaos in interacting systems, which use quantum chaos indicators to pinpoint an ergodicity breaking transition.
In most cases of interest for current studies, the critical parameters of the transition are usually not known in advance.
A typical example is the random-field spin-1/2 Heisenberg chain, which has recently experienced a renewed interest to unveil the eventual breakdown of quantum chaos~\cite{suntajs_bonca_20, suntajs_bonca_20a, sels_polkovnikov_21, leblond_sels_21, sels_polkovnikov_22, kieferemmanouilidis_unanyan_20, luitz_barlev_20, kieferemmanouilidis_unanyan_21, panda_scardicchio_20, sierant_delande_20, sierant_lewenstein_20, abanin_bardarson_21, corps_molina_21, hopjan_orso_21, crowley_chandran_22, vidmar_krajewski_21, sels_22, morningstar_colmenarez_22, sierant_zakrzewski_22, krajewski_vidmar_22}.
Here we study a single-particle problem, which is assumed to be more accessible to numerical methods and hence a precise determination of the onset of localization is expected to be feasible.
A specific question we address is the following: provided that one is not aware of previous work on Anderson localization in 2D, is it possible to conclude, based on calculations of quantum chaos indicators in finite systems, that the 2D Anderson model localizes at any nonzero disorder in the thermodynamic limit? As we show here, the answer is indeed affirmative.
Our central result are the scaling solutions of quantum chaos indicators as functions of an effective parameter $u = (W \ln V)^{-1}$, where $V$ is the number of lattice sites, which also equals the single-particle Hilbert space dimension.
In particular, upon identifying the parameter $u$, the results follow their asymptotic form already at rather small system sizes.
These results hence provide a positive example about the ability of exact numerical approaches to accurately describe the fate of a quantum-chaos to localization transition in the thermodynamic limit.

\subsection{Anderson model}

The focus of this paper is the 2D Anderson model on a square lattice $L\times L$,
\begin{align} \label{def_Ham}
 \hat H & = -t \sum_{\langle i,j\rangle}  \left( \hat c_i^\dagger \hat c_j + {\rm h.c.} \right) + \frac{W}{2} \sum_{i=1}^{V} w_{i} \hat n_i  \, ,
\end{align}
assuming periodic boundary conditions. Here, $\hat c_i^\dagger$ and $\hat c_i$ are the spinless fermion creation and annihilation operators, respectively, at site $i$, and $\hat n_i = \hat c_i^\dagger \hat c_i$. The number of lattice sites is $V=L^2$, which also corresponds to the single-particle Hilbert space dimension. 
In the first term in Eq.~(\ref{def_Ham}), describing the nearest neighbour hopping, we set the corresponding kinetic energy scale $t\equiv 1$ throughout our calculations. Representing disorder in the system, random on-site potentials $w_j$ in the second term in Eq.~(\ref{def_Ham}) are independent and identically distributed random numbers $w_j \in [-1, 1],$ drawn from a uniform distribution. The degree of the disorder is hence controlled by the disorder strength parameter $W$.

The Anderson model in Eq.~(\ref{def_Ham}) is a paradigmatic model to study Anderson localization and has been a focus of research over nearly 5 decades, see, e.g., Refs.~\cite{kramer_mackinnon_93, abrahams201050, RevModPhys.80.1355, markos_06, scardicchio_thiery_17, suntajs_prosen_21} for reviews.
Below we highlight several properties of the Anderson model that are relevant for our study.

Perhaps the most widely studied Anderson model is the one defined on a cubic (3D) lattice, for which the existence of the transition at nonzero disorder is well established~\cite{abrahams_anderson_79, mackinnon_kramer_81, kramer_mackinnon_93}.
However, even in this case, as noted in the Anderson's Nobel lecture~\cite{anderson_78}, one needs to resort to numerical calculations to pinpoint the actual value of the transition point.
The latter has indeed been determined to high accuracy using various numerical techniques in finite systems~\cite{slevin_ohtsuki_18, suntajs_prosen_21}, and nowadays the established transition point in the thermodynamic limit is $W_{\rm c} \approx 16.5$.

We note that very useful indicators to accurately pinpoint the transition point in the 3D Anderson model are based on quantum chaos indicators, which are discussed in more detail in Sec.~\ref{sec:indicators}.
A paradigmatic example of the latter are spectral statistics, for which a connection between the 3D Anderson model below the transition (i.e., at $W < W_{\rm c}$) and the predictions of the random matrix theory (RMT) were made already in the mid-eighties~\cite{altshuler_shklovskii_86}, shortly after the formulation of the quantum chaos conjecture~\cite{casati_valzgris_80, bohigas_giannoni_84}.
The transition in finite systems then corresponds to a crossover between a quantum chaotic regime at $W \ll W_{\rm c}$, in which spectral statistics follow predictions from the Gaussian orthogonal ensemble (GOE), and a localized regime at $W \gg W_{\rm c}$, in which spectral statistics are Poissonian~\cite{shklovskii_shapiro_93, hofstetter_schreiber_93, tarquini_biroli_17, suntajs_prosen_21, sierant_lewenstein_22, garciamata_martin_22}.

\begin{figure*}[!t]
\centering
\includegraphics[width=2.00\columnwidth]{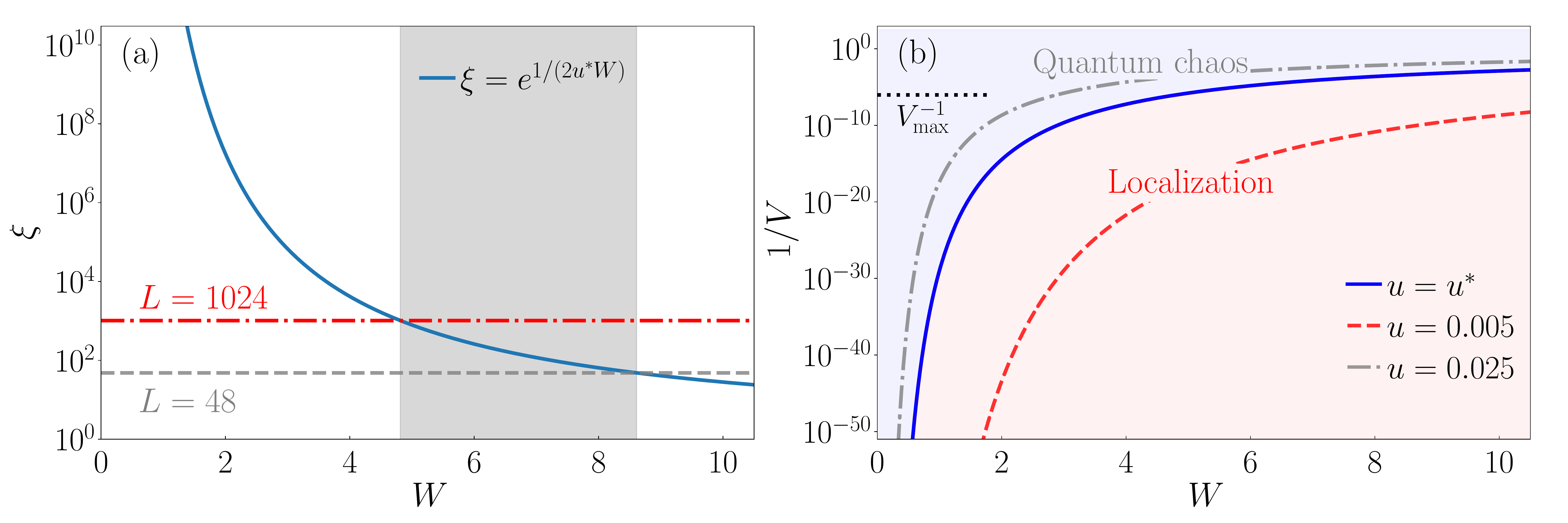}
\caption{
(a) Localization length $\xi$ from Eq.~(\ref{eq:def_xi}) versus disorder $W$.
We set $a=1/(2 u^*)$, see Eq.~(\ref{def_a}), where $u^* = 0.015$, and $b=1$.
The dashed and dashed-dotted horizontal lines denote the smallest ($L=48$) and largest ($L=1024$) linear system size $L = \sqrt{V}$ used in our calculations, respectively. The shaded region marks the regime in which $\xi = L$ for the system sizes under investigation.
(b) Flow of $1/V$ versus $W$.
Solid line denotes the flow of the scale-invariant point set by $u=u^*=0.015$, see Eq.~\eqref{def_u_star}.
The dashed and dashed-dotted lines denote the flow of the effective parameter $u$ from Eq.~(\ref{def_u}) in localized and quantum chaotic regimes at $u=0.005 < u^*$ and $u=0.025 > u^*$, respectively.
Horizontal dotted line denotes the inverse maximal system size studied here, $V_{\rm max}^{-1} = 2^{-20} \approx 10^{-6}$.
}
\label{fig0}
\end{figure*}

\subsection{Single-parameter scaling (SPS) hypothesis}

Resolving the fate of the Anderson transition in 2D turned out to be a much more difficult task compared to 3D.
Before the introduction of the single-parameter scaling (SPS) hypothesis~\cite{abrahams_anderson_79}, many studies were interpreted in terms of a transition occurring at nonzero disorder, i.e., $W_{\rm c}>0$~\cite{edwards_thouless_72, licciardello_economou_75, licciardello_thouless_75, licciardello_thouless_75b, yoshino_okazaki_77, weaire_srivastava_77, prelovvsek1978diffusion, PhysRevLett.40.1596, PhysRevLett.42.1492, stein_krey_80, kramer1981numerical}.
According to the SPS hypothesis of localization~\cite{abrahams_anderson_79}, which builds on  the assumption that the real-space wave functions in the localized regime decay exponentially with distance, the 2D system exhibits Anderson localization at any nonzero $W$ in the thermodynamic limit.
Soon after the introduction of the SPS hypothesis, numerical studies confirmed its relevance at moderate and large disorder~\cite{mackinnon_kramer_81, mackinnon1983scaling}, and identified exponential divergence of the localization length of the form~\cite{mackinnon_kramer_81},
\begin{equation}\label{eq:def_xi}
    \xi = b \, e^{\frac{a}{W}}\;,
\end{equation}
where $a$ is a constant and $b$ is a parameter that may eventually include subleading contributions to $W$-dependence.
At weak disorder, perturbative results established logarithmic finite-size effects that suggested absence of a true metallic behavior, and hence this regime is also referred to as the regime of weak localization~\cite{kramer_mackinnon_93}.
Finite-size effects that scale as logarithms of system volume $V$ render the numerical analysis of the problem notoriously difficult and make the true nature of the transition hard to address.
For example, some early numerical works suggested evidence against the SPS hypothesis and highlighted a sub-exponential decay of wave functions at weak disorder~\cite{Kaveh_1981, pichard_sarma_81a, pichard_sarma_81b, Schreiber_1985}.

In the actual numerical studies of finite 2D systems, the origin of difficulties to detect localization can be understood as being a consequence of $\xi > L$, i.e., the localization length $\xi$ from Eq.~(\ref{eq:def_xi}) exceeding the linear system size $L = \sqrt{V}$, see Fig.~\ref{fig0}(a).
While previous studies of the 2D Anderson model mostly focused on properties of conductance and the corresponding distributions~\cite{slevin_asada_04, prior_somoza_05, somoza_prior_09, somoza_ledoussal_15, lopezbezanilla_froufeperes_18}, here we focus on quantum chaos indicators that imprint properties of Hamiltonian spectrum, eigenfunctions, and matrix elements of observables.
We show that the equality $\xi = L$, which is fulfilled at some intermediate disorder $W^*$ for system sizes under consideration, gives rise to a quantum-chaos to localization crossover at roughly the same $W^*$.
Moreover, $W^*$ is a function of the system size $V$, which we denote as $W^* = W^*(V)$.
For the available system sizes $V \lesssim 10^6$ one typically expects $W^*(V) \gtrsim 5$~\cite{edwards_thouless_72, licciardello_economou_75, licciardello_thouless_75, licciardello_thouless_75b, yoshino_okazaki_77, weaire_srivastava_77, prelovvsek1978diffusion, PhysRevLett.40.1596, PhysRevLett.42.1492, kramer1981numerical, mackinnon_kramer_81, mackinnon1983scaling, schreiber_ottomeier_92, slevin_asada_04}.
We observe that the regime $W < W^*(V)$ exhibits extremely robust signatures of quantum chaos.
The latter are sometimes so robust that the convergence to RMT predictions even improves by increasing the system size $V$.
This may easily be interpreted as evidence of a true localization transition occurring at $W_{\rm c} > 0$ in the thermodynamic limit.

Nevertheless, as will be demonstrated in Sec.~\ref{sec:indicators}, our results based on state-of-the-art numerical approaches are in accordance with the SPS hypothesis and suggest $W_{\rm c} = 0$, i.e., the absence of localization transition at nonzero disorder.
The results can be summarized in two steps.

The first step relies on numerical observations.
Monitoring the departure of various quantum chaos indicators (to be introduced in Sec.~\ref{sec:indicators}) from the RMT predictions, we observe a systematic drift of the crossover point $W^*(V) \to 0$ upon increasing $V$.
Specifically, we observe a drift of $1/W^*(V)$ that is logarithmic in $V$.
This motivates us to introduce the effective parameter
\begin{equation} \label{def_u}
    u = \left(W\ln V\right)^{-1} \;.
\end{equation}
Remarkably, when studying quantum chaos indicators as a function of $u$, we observe a scale-invariant crossover point $u^*$ that emerges for all indicators under investigation at approximately the same value,
\begin{equation} \label{def_u_star}
u^* = \left[W^*(V) \, \ln V\right]^{-1}\approx 0.015 \;.
\end{equation}
The flow of the disorder at the scale-invariant point $W^*$ with $V$ is demonstrated in Fig.~\ref{fig0}(b).
We refer to the regime $u > u^*$ [i.e., $W < W^*(V)$] as {\it prelocalization}, thereby highlighting that while a given system appears to be quantum chaotic, it localizes in the thermodynamic limit $V \to \infty$.

In the second step, we test the dependence of various quantum chaos indicators as a function of $\xi/L$, assuming the parameter $b$ in Eq.~(\ref{eq:def_xi}) to be a constant.
We find excellent scaling collapses of the results in a broad interval of system sizes $10^3 \lesssim V \lesssim 10^6$, thereby confirming validity of the SPS hypothesis in the 2D Anderson model.

The central question is then to establish the connection between the numerical observations in the first step and the SPS property in the second step.
Using Eq.~(\ref{eq:def_xi}) and setting $L = \sqrt{V}$, one can express
\begin{align}
    \ln(\xi/L) & = \frac{a}{W} - \frac{1}{2} \ln V + \ln b \nonumber \\
    & = \left( a \, u - \frac{1}{2} \right) \ln V + \ln b \;, \label{def_xi_L}
\end{align}
where in the last step we introduced the parameter $u$ from Eq.~(\ref{def_u}).
Equation~(\ref{def_xi_L}) offers two interesting insights.
The first is that the ratio $\xi/L$, which depends on both $W$ and $V$, can be conveniently expressed as a function of two parameters, the effective parameter $u$ from Eq.~(\ref{def_u}) and the logarithm of the system size $V$.
The second insight is that the scale invariance of $\xi/L$ gives rise to the condition
\begin{equation} \label{def_a}
    a \, u^* - \frac{1}{2} = 0 \;\;\; \longrightarrow \;\;\; a = \frac{1}{2 u^*} \;.
\end{equation}
This allows us to simplify Eq.~(\ref{def_xi_L}) as
\begin{equation}\label{eq:def_xi_l}
\ln \left(\xi/L\right) = \left(\frac{u - u^*}{u^*}\right)\ln L + {\rm const} \;,
\end{equation}
where the constant equals $\ln b$.

\begin{figure*}[!t]
\centering
\includegraphics[width=2.00\columnwidth]{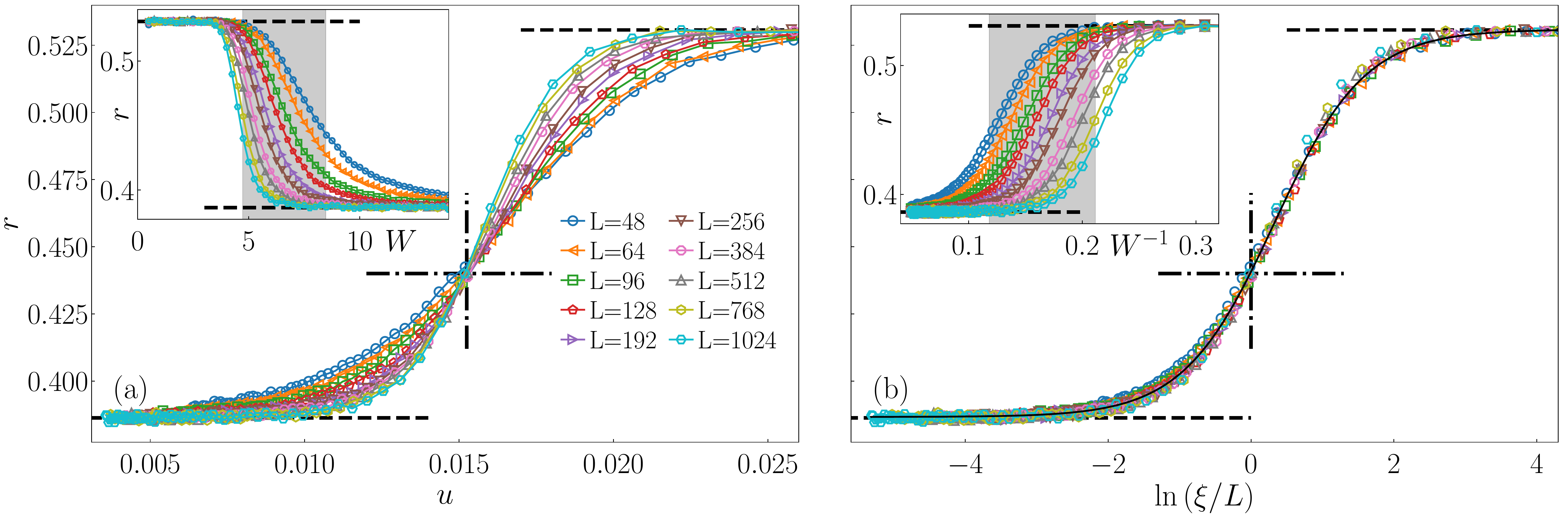}
\caption{
Study of the average level spacing ratio $r$.
Results are shown for different linear system sizes $L=\sqrt{V}$, as indicated in the legend.
(a) Behavior of $r$ as a function of disorder $W$ in the inset, and as a function of the effective parameter $u$, see Eq.~(\ref{def_u}).
The curves in the main panel intersect at the scale-invariant point $u^* = \left(W^*\ln V\right)^{-1} \approx 0.0153,$
indicated by the vertical dashed-dotted line, while the horizontal dashed-dotted line denotes the corresponding value of $r$ at this point, $r\approx 0.44$.
(b) Behavior of $r$ as a function of inverse disorder $W^{-1}$ in the inset, and as a function of $\ln(\xi/L)$ in the main panel using $\xi = \exp\left(1/(2u^*W)\right)$, see Eqs.~(\ref{eq:def_xi}) and~(\ref{def_a}).
Bottom and top dashed horizontal lines in all panels denote the limiting values in the GOE and Poisson regimes, $r_\mathrm{GOE}\approx 0.5307$ and $r_\mathrm{P}\approx 0.3863,$ respectively. 
The solid line in the main panel is the function
$\bar r + \Delta r \tanh[k(\ln(\xi/L)-c_0)]$, where $\bar r = (r_{\rm GOE} + r_{\rm P})/2= 0.46$ and $\Delta r = (r_{\rm GOE} - r_{\rm P})/2 = 0.07$, and the fitting parameters are $k=0.79$ and $c_0 = 0.32$.
The shaded regions in both insets are identical to the one in Fig.~\ref{fig0}(a).
}
\label{fig2}
\end{figure*}

The remainder of the paper is organized as follows.
We identify the scale-invariant point $u^*$ and test validity of the SPS hypothesis for three quantum chaos indicators:
the average gap ratio in Sec.~\ref{sec:spectral}, the ratio of variances of observable matrix elements in Sec.~\ref{sec:matele}, and the entanglement entropy of many-body eigenstates in Sec.~\ref{sec:eigenstates}.
Remarkably, all quantum chaos indicators exhibit even quantitatively very similar results, and provide strong support in favor of the SPS hypothesis and $W_{\rm c} = 0$ in the thermodynamic limit.
We complement these results by two other results.
We observe in Sec.~\ref{sec:spectral} that the spectral form factor $K(\tau)$ [to be defined in Eq.~(\ref{eq:def_SFF})] exhibits a broad plateau in the prelocalized regime $u > u^*$, and in Sec.~\ref{sec:eigenstates} we observe that the derivative of the participation entropy [to be defined in Eq.~(\ref{def:Sp})] w.r.t.~disorder, $d\overline{S_{\rm P}}/dW$, exhibits a peak in the vicinity of the scale-invariant point $u^*$.
Both results should be contrasted to the behavior in the 3D Anderson model, for which it was recently shown that at the transition $K(\tau)$ exhibits a broad scale-invariant plateau, and $d\overline{S_{\rm P}}/dW$ exhibits a peak that eventually corresponds to a divergence in the thermodynamic limit~\cite{suntajs_prosen_21}.

\section{Indicators of quantum chaos} \label{sec:indicators}

From now on we focus on various quantum chaos indicators based on spectral properties of single-particle eigenenergies (Sec.~\ref{sec:spectral}), matrix elements of observables in single-particle Hamiltonian eigenstates (Sec.~\ref{sec:matele}), and the wavefunction structure of both single-particle and many-body eigenstates (Sec.~\ref{sec:eigenstates}).
We numerically obtain exact single-particle eigenstates of the 2D Anderson model~(\ref{def_Ham}) using either full diagonalization or shift-invert method~\cite{pietracaprina2018shift}.
Using the latter, we obtained 500 single-particle eigenstates from the middle of the spectrum for Hamiltonian matrices of dimension $V \times V$, with the largest $V_{\rm max}=L_{\rm max}^2=2^{20} = 1048576$.
When applicable, we compare the results to the corresponding predictions of the Gaussian orthogonal ensemble (GOE) of the RMT.

Since the Anderson model in Eq.~(\ref{def_Ham}) is quadratic, agreement of quantum chaos indicators with GOE predictions suggests that the model is quantum-chaotic quadratic~\cite{lydzba_rigol_21, PhysRevB.104.214203}, and hence one observes single-particle quantum chaos.
Note that the latter does not necessary exhibit identical properties as interacting systems for which one encounters many-body quantum chaos~\cite{dalessio_kafri_16}.
For example, it was recently shown that the spectral statistics of {\it many-body} eigenenergies of quantum-chaotic quadratic Hamiltonians does not comply with GOE predictions~\cite{lydzba_rigol_21} (see also~\cite{liao_vikram_20, winer_jian_20}), the entanglement entropy of the corresponding many-body eigenstates is not necessary maximal~\cite{lydzba_rigol_vidmar_prl_2021_PhysRevLett.125.180604, lydzba_rigol_21, PRXQuantum.3.030201, PhysRevE.106.034118}, and the matrix elements of observables in many-body eigenstates do not exhibit eigenstate thermalization~\cite{lydzba_mierzejewski_23}.

\begin{figure*}[!t]
\centering
\includegraphics[width=2.00\columnwidth]{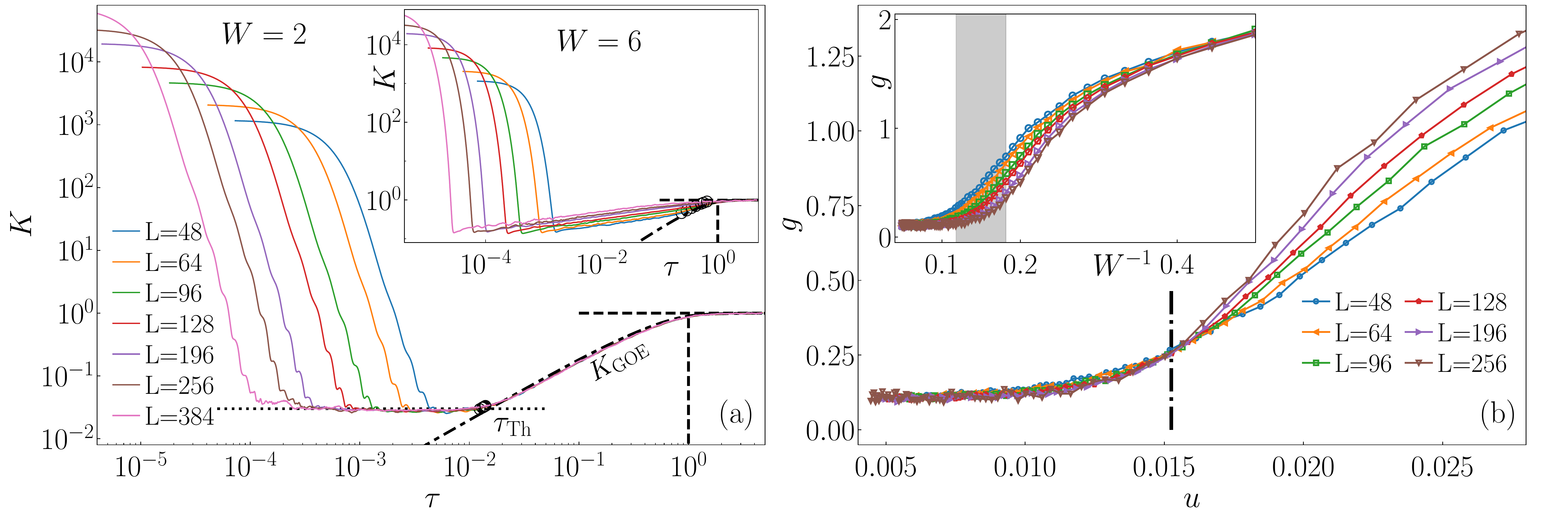}
\caption{(a)
The spectral form factor $K(\tau)$ in the 2D Anderson model for different linear system sizes $L=\sqrt{V}$, as indicated in the legend.
Dashed-dotted lines denote the GOE result $K_{\rm GOE}(\tau) = 2\tau - \tau\ln(1+2\tau),$ open circles denote the extracted values of the scaled Thouless time $\tau_{\rm Th} $ and the vertical dashed line is the scaled Heisenberg time $\tau_{\rm H}=1.$
Results are shown at $W=2$ in the main panel and $W=6$ in the inset.
The horizontal dotted line in the main panel corresponds to $K\approx 0.03$.
(b) Behavior of the quantum chaos indicator $g$ from Eq.~(\ref{def_g}), which quantifies the ratio of the Heisenberg and Thouless times.
Results are shown as a function of inverse disorder $W^{-1}$ in the inset and as a function of the effective parameter $u$ in the main panel.
The vertical dashed-dotted line in the main panel corresponds to $u^* = 0.0153$ extracted from Fig.~\ref{fig2}, and the shaded area in the inset corresponds to the same interval of disorders $W^*$ as in the insets of Fig.~\ref{fig2}.
}
\label{fig1}
\end{figure*} 

\subsection{Spectral analysis} \label{sec:spectral}

We first study the average level spacing ratio $r$~\cite{oganesyan_huse_07}.
It is defined through the level spacing ratio
\begin{equation}
r_\alpha = \min\{\delta_\alpha, \delta_{\alpha-1}\}/\max\{\delta_\alpha, \delta_{\alpha-1}\}\;,
\end{equation}
where $\delta_\alpha = E_{\alpha+1} - E_{\alpha}$ is the level spacing between the energy-ordered single-particle eigenstates $|\alpha\rangle$ and $|\alpha+1\rangle$ of the Hamiltonian~(\ref{def_Ham}).
We then obtain $r=\langle \langle r_\alpha \rangle_\alpha \rangle$ by the average $\langle ... \rangle_\alpha$ over eigenstates around the center of the spectrum and the average $\langle ... \rangle$ over different realizations of the disorder distributions.
The asymptotic values are $r \approx r_{\rm GOE} \approx 0.5307$ in the chaotic regime~\cite{atas_bogomolny_13} and $r \to r_{\rm P} = 2\ln(2)-1 \approx 0.3863$ in the localized regime~\cite{oganesyan_huse_07}.

The inset of Fig.~\ref{fig2}(a) shows the raw results for $r(W)$ at different $V$.
They exhibit a broad crossover regime at $4 \lesssim W \lesssim 10$ and a seemingly very robust GOE regime at weak disorder $W \lesssim 4$.
From this result only it is not possible to convincingly argue about the existence of localization at any nonzero disorder in the thermodynamic limit $V \to \infty$.

Robustness of localization becomes more apparent when $r$ is plotted as a function of $1/W$, as shown in the inset of Fig.~\ref{fig2}(b).
These results exhibit a linear drift of the crossover region with the logarithm of the lattice volume $V$.
This motivates us to introduce a scaled parameter $u=1/(W \ln V)$ from Eq.~(\ref{def_u}).
Results for $r$ versus $u$ are shown in the main panel of Fig.~\ref{fig2}(a).
They exhibit a scale-invariant point at $u^* = 0.0153$.
Emergence of a scale invariant point in $r(u)$ suggests that the localized regime described by the Poisson statistics extends over all positive values of $W$ in the limit $V \to \infty$.

In Fig.~\ref{fig2}(b) we then test the SPS hypothesis by plotting $r$ as a function of $\ln(\xi/L)$, where $\xi = b e^{1/(2u^* W)}$ is given by Eqs.~(\ref{eq:def_xi}) and~(\ref{def_a}).
Since the parameter $b$ only determines a global horizontal shift of the scaled results, we set $b = 1$ further on.
We observe a remarkably good scaling collapse of the data for essentially all system sizes under consideration.

In passing, we note that the definition of the localization length $\xi = e^{1/(2u^* W)}$ has a very simple interpretation:
it corresponds to the linear systems size $L$ at the scale-invariant point, i.e., $\xi = L$ at the point when $W$ equals $W^*(V) = 1/(u^* \ln V)$, see also Eq.~(\ref{def_u_star}).
The dependence of $\xi$ on $W$ is shown in Fig.~\ref{fig0}(a).

Next, we study the \emph{spectral form factor} (SFF) $K(\tau),$ which is the Fourier transform of the two-point spectral correlation function. It is defined as
\begin{equation}\label{eq:def_SFF}
    K(\tau) = \frac{1}{Z}\left\langle \left|\sum\limits_{n=1}^V f(\varepsilon_n)\ee^{-\ii 2\pi\varepsilon_n\tau}\right|^2\right\rangle.
\end{equation}
Here, $\{\varepsilon_1 \leq \varepsilon_2 \leq \dots, \varepsilon_V\}$ is the complete set of Hamiltonian single-particle eigenvalues after spectral unfolding, which sets the mean level spacing to unity. We thus refer to $\tau$ as the scaled time and the averaging $\langle \dots \rangle$ is performed over different disorder realizations of $\hat{H}$ in Eq.~(\ref{def_Ham}).
Details of the filtering function $f(\varepsilon)$ [used to minimize the finite-size effects], normalization $Z$ and the unfolding procedure are described in Ref.~\cite{suntajs_bonca_20a} and Appendix~\ref{sec:app_sff}.

We explore the scaling of two relevant time scales, the Heisenberg and the Thouless time.
The first is proportional to the inverse single-particle mean level spacing and corresponds to the longest time scale under investigation.
In scaled units, the Heisenberg time is set to unity, $\tau_{\rm H} = 1.$
The Thouless time is defined as the onset time of quantum chaos, i.e., in scaled units $\tau_{\rm Th}$ is the time after which $K(\tau)$ in chaotic systems becomes universal and well described by the GOE prediction~\cite{mehta_91}.
In contrast, $K(\tau \ll 1) \approx 1$ in strongly non-chaotic regime characterized by Poissonian level statistics.
In this case, therefore, the SFF $K(\tau)$ does not represent a tool that accurately determines $\tau_{\rm Th}$, but it merely yields $\tau_{\rm Th} \approx \tau_{\rm H}$.

Recently, further universal features of the $K(\tau)$ at $\tau \leq \tau_{\rm Th}$ have been observed at the transition point of the 3D Anderson model~\cite{suntajs_prosen_21}.
There, a scale-invariant $K(\tau) \approx \mathrm{const} < 1$ emerges for a broad interval of $\tau$.
Since the outlined behaviour is only limited to the transition point, one can consider independence of $\tau_{\rm Th}$ on $L$ as a criterion for the transition~\cite{suntajs_prosen_21, sierant_delande_20}.
Intriguingly, very similar behavior of $K(\tau)$ was recently observed at the ergodicity breaking phase transition in the interacting avalanche model~\cite{PhysRevLett.129.060602}.
These observations set a question whether similar universal features can be observed in the 2D Anderson model, and, more generally, to which extent one can use the SFF $K(\tau)$ to infer the fate of the quantum-chaos to localization transition in the thermodynamic limit.

Results shown in Fig.~\ref{fig1}(a) show quite unexpected phenomenology.
While the $K(\tau)$ at $W=6$, see the inset of Fig.~\ref{fig1}(a), exhibits a flow of the Thouless time $\tau_{\rm Th} \to 1$ and is hence consistent with restoration of localization, the result at $W=2$, see the main panel of Fig.~\ref{fig1}(a), is not.
It exhibits two intriguing features:
a seemingly scale-invariant value of $\tau_{\rm Th} = {\rm const} < 1$,
and a broad, scale-invariant plateau of $K(\tau)$ that extends over at least two orders of magnitude.
In Fig.~\ref{Supp1} in Appendix~\ref{sec:app_sff}, we show that a very similar structure of $K(\tau)$ is observed in the broad prelocalized regime at $W \ll W^*(V)$, where $W^*(V) \approx 5$ for the largest accessible system sizes.

We note that in physical units, the Thouless time $t_{\rm Th}$ is calculated as $t_{\rm Th} = \tau_{\rm Th} t_{\rm H}$~\cite{suntajs_bonca_20, suntajs_prosen_21}, where the Heisenberg time $t_{\rm H}$ is proportional to $L^2$, i.e., the inverse mean level spacing in a 2D system.
Hence, the observation of $\tau_{\rm Th} = {\rm const}$ at $W \ll W^*(V)$ implies a diffusive scaling $t_{\rm Th} \propto L^2$.
Such a scaling is understood as a finite-size effect since it occurs in systems for which $L < \xi$, see Fig.~\ref{fig0}(a).

Results for $K(\tau)$ at weak disorder $W < W^*(V)$, and their comparison to the results at the transition point in the 3D Anderson model, may suggest robustness of quantum chaos associated with a universal structure of $K(\tau)$ at all times.
However, as discussed below, this quantum-chaotic regime is likely a finite-size feature that evolves towards localization in the thermodynamic limit.

To elucidate the scaling properties of the Thouless time, we analyse the logarithmic ratio of the Heisenberg and Thouless time~\cite{suntajs_bonca_20a},
\begin{equation} \label{def_g}
    g = \log_{10}\left(\tau_{\rm H}/\tau_{\rm Th}\right) = -\log_{10} \tau_{\rm Th}\;,
\end{equation}
and study its behavior at different $W$ and $V$.
Recently, $g$ has been used as a reliable indicator to pinpoint the transition in the 3D Anderson model~\cite{suntajs_prosen_21, sierant_delande_20} and the interacting avalanche model~\cite{PhysRevLett.129.060602}. As argued in Ref.~\cite{suntajs_bonca_20a}, the indicator $g$ interpolates between the quantum chaotic regime, $g \to \infty,$ and the nonchaotic regime, $g \to 0,$ in the thermodynamic limit. In the 3D Anderson model, the scale-invariant point at which $g^*$ is independent of $L$ reliably pinpoints the transition~\cite{sierant_delande_20, suntajs_prosen_21}.

As shown in the inset of Fig.~\ref{fig1}(b), $g$ versus $W^{-1}$ in the 2D Anderson model does actually not exhibit any scale-invariant {\it point}.
Instead, it exhibits seemingly two broad scale-invariant {\it regimes} that emerge at large and small $W$.
This property may already be interpreted as a signature of the absence of localization transition at nonzero $W$.

Even more importantly, we observe a drift of the rising point of $g(W^{-1})$ towards smaller $W$, which is indicated by the shaded region in the inset of Fig.~\ref{fig1}(b) and it roughly occurs at $W \approx W^*(V)$.
In the scenario of absence of localization transition at nonzero $W$, this drift is expected to persist in the thermodynamic limit.
This expectation is confirmed in the main panel of Fig.~\ref{fig1}(b), in which we plot $g$ versus the effective parameter $u$.
We observe a scale-invariant regime with $g \ll 1$ at $u < u^*$, which indicates localization, and a nonuniversal regime with diverging $g$ at $u > u^*$.
Since the fixed point $u^*$ implies the flow of the disorder $W^*(V) \to 0$ in the thermodynamic limit $V \to \infty$, this implies that the quantum-chaotic regime vanishes with increasing $V$.
This interpretation is consistent with the results for the average level spacing ratio $r$ in Fig.~\ref{fig2}.

\subsection{Matrix elements of observables} \label{sec:matele}

To complement the studies of the spectral statistics, we next investigate the properties of matrix elements of local observables in single-particle energy eigenstates. 
The relevant point of comparison is the single-particle eigenstate thermalization~\cite{lydzba_rigol_21}, which is expected to describe properties of these matrix elements in quantum-chaotic quadratic Hamiltonians.

We note that the single-particle eigenstate thermalization carries many similarities, but also important differences, with respect to the eigenstate thermalization hypothesis (ETH) that is applied to matrix elements in {\it many-body} eigenstates of quantum-chaotic interacting systems~\cite{deutsch_91, srednicki_94, srednicki_99, rigol_dunjko_08, dalessio_kafri_16, mori_ikeda_18, deutsch_18}.
While the ETH is relatively well established and has been a subject of extensive numerical studies~\cite{santos2010_1, rigol_dunjko_08, pop1}, the concept of single-particle eigenstate thermalization has not been studied until recently~\cite{lydzba_rigol_21, PhysRevE.106.034118}.
In particular, we are not aware of any previous studies that use the comparison to single-particle eigenstate thermalization as an indicator of quantum-chaos to localization transition.

\begin{figure}[!]
\centering
\includegraphics[width=1.00\columnwidth]{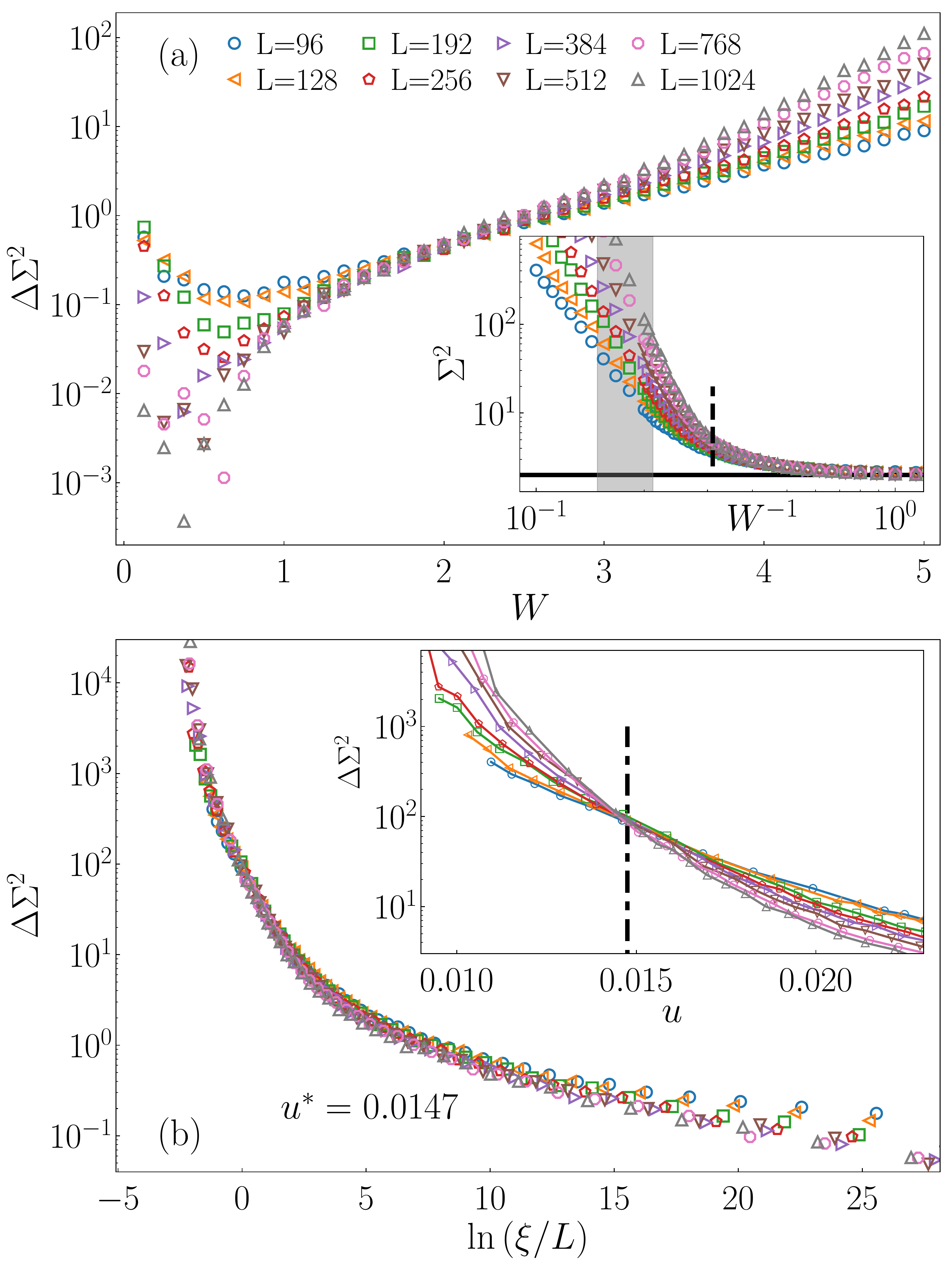}
\caption{
Ratio of variances of matrix elements of the site occupation operator $\hat{n}_i$ at the center of the lattice, $i=(L/2, L/2).$
Results are shown for different linear system sizes $L=\sqrt{V}$, as indicated in the legend.
(a) $\Delta\Sigma^2$, see Eq.~(\ref{eq:def_sigma_diff}), as a function of disorder $W$ in the main panel, and $\Sigma^2$, see Eq.~(\ref{eq:def_ratio_vars}), as a function of inverse disorder $W^{-1}$ in the inset.
The horizontal line in the inset denotes $\Sigma^2_{\rm GOE} = 2$, while the vertical dashed line denotes the disorder value at which $r$ in Fig.~\ref{fig2}(a) still displays the GOE behavior for the largest studied system.
(b) $\Delta\Sigma^2$ as a function of the effective parameter $u$ in the inset, and $\Delta\Sigma^2$ as a function of $\ln\left(\xi/L\right)$ in the main panel.
We extract the scale-invariant point $u^* = 0.0147$ in the inset, see the vertical dashed-dotted line, and use it in the definition of $\xi = \exp(1/(2u^*W))$ in the main panel.
}
\label{fig3}
\end{figure} 

A specific quantity that we study is the ratio $\Sigma^2$ of variances of the diagonal and off-diagonal matrix elements, to be defined in Eq.~(\ref{eq:def_ratio_vars}).
In quadratic systems that obey single-particle eigenstate thermalization, as well as in interacting systems that obey the ETH, one expects $\Sigma^2 \to \Sigma^2_{\rm GOE} = 2$, if the system belongs to the GOE universality~\cite{dalessio_kafri_16}.
Indeed, this has been observed both in quantum-chaotic interacting~\cite{mondaini_fratus_16, jansen_stolpp_19, richter_dymarsky_20, schoenle_jansen_21} and quantum-chaotic quadratic Hamiltonians~\cite{PhysRevB.104.214203, PhysRevE.106.034118}.

For a given observable $\hat{O},$ we first calculate the variance of its diagonal matrix elements  $O_{\alpha \alpha} \equiv \langle \alpha| \hat O |\alpha\rangle$ in the single-particle energy eigenstates $\ket{\alpha}$ as
\begin{equation}\label{eq:def_var_diag}
\sigma^2_{\rm diag} = \vert \mu \vert^{-1} \sum\limits_{\ket{\alpha}\in \mu} O_{\alpha \alpha}^2 - \left(\vert \mu \vert^{-1} \sum\limits_{\ket{\alpha}\in \mu} O_{\alpha \alpha}\right)^2.
\end{equation}
Here, $\mu$ is a set of 500 eigenstates $(\vert \mu\vert=500)$ centered around the mean energy of the single-particle spectrum. 
As highlighted in~\cite{mondaini_fratus_16, jansen_stolpp_19}, the number of eigenstates included in the average relative to the total number of states should be a small number that quickly vanishes when increasing the system size.
Similarly, we calculate the off-diagonal matrix elements $O_{\alpha \beta} \equiv \langle \alpha| \hat O |\beta\rangle$ and obtain their variance as
\begin{equation}\label{eq:def_var_offdiag}
\sigma^2_{\rm off} = \vert \mu' \vert^{-1} \sum\limits_{\substack{\ket{\alpha}, \ket{\beta} \in \mu \\ \ket{\alpha} \ne \ket{\beta}}} O_{\alpha \beta}^2 
- \left(\vert \mu' \vert^{-1} \sum\limits_{\substack{\ket{\alpha}, \ket{\beta} \in \mu \\ \ket{\alpha} \ne \ket{\beta}}} O_{\alpha \beta}\right)^2,
\end{equation}
where $\vert \mu' \vert$ is the number of the off-diagonal matrix elements, $\vert \mu' \vert = \vert \mu \vert^2 - \vert \mu \vert.$ For each disorder realization, we  calculate the ratio of variances of diagonal and off-diagonal matrix elements and then average the results over different disorder realizations, i.e., we define the ratio of variances as
\begin{equation}\label{eq:def_ratio_vars}
    \Sigma^2 = \left\langle \frac{\sigma^2_{\rm diag}}{\sigma^2_{\rm off}} \right\rangle \;,
\end{equation}
where $\langle ... \rangle$ denotes the averaging over different Hamltonian realizations.
For convenience, we typically study the difference between $\Sigma^2$ and the GOE value $\Sigma^2_{\rm GOE}=2$,
\begin{equation}\label{eq:def_sigma_diff}
    \Delta\Sigma^2 = \left|\Sigma^2 - \Sigma^2_{\rm GOE} \right| \;.
\end{equation}
As an observable we consider the site occupation operator $\hat{n}_i = \hat{c}^\dagger_i \hat{c}_i$ at the center of the lattice, i.e., at the site $i=(L/2, L/2)$.
We note that we have also tested the implementation of $\Sigma^2$ where in Eq.~(\ref{eq:def_ratio_vars}) the ratio of mean variances is taken rather than the mean of the ratios. While this choice does not seem to particularly affect the scaling in the GOE regime, it appears less stable at larger disorders and the results exhibit larger fluctuations.

Results for $\Delta\Sigma^2$ as a function of disorder $W$ are shown for different system sizes in Fig.~\ref{fig3}.
We first note a remarkably good agreement of the numerical results with the GOE prediction $\Delta\Sigma^2=0$ in the regime $W \leq 1$.
In particular, we find $\Delta\Sigma^2 \lesssim 10^{-2}$ for the largest system with $L=1024$, which is comparable with the most accurate studies carried out so far in interacting systems~\cite{schoenle_jansen_21}.
Increasing the disorder, we observe a gradual increase of $\Delta\Sigma^2$ in accordance with the onset of localized behavior. Interestingly, while spectral statistics for the largest available system can still show $r=r_{\rm GOE}$ in the crossover regime, see the vertical dashed line in the inset of Fig.~\ref{fig3}(a), $\Sigma^2$ is already far from the GOE prediction.

We test the predicting power of $\Delta\Sigma^2$ to pinpoint the localization transition by plotting $\Delta\Sigma^2$ as a function of the effective parameter $u$ in the inset of Fig.~\ref{fig3}(b).
Quite surprisingly, we observe a scale-invariant crossover point that is quantitatively very close to the value $u^* = 0.0153$ obtained in the study of the average level spacing ratio in Sec.~\ref{sec:spectral}.
Note, however, that the value of $\Delta\Sigma^2$ at the scale-invariant point is rather large, $\Delta\Sigma^2 \approx 10^2$.
Moreover, we again cast the results as a function of $\ln\left(\xi/L\right)$, see the main panel of Fig.~\ref{fig3}(b), and obtain an excellent data collapse for different system sizes.
These results suggest, quite non-trivially, that a proper indicator of the statistical properties of matrix elements, such as $\Delta\Sigma^2$ studied here, can detect both the absence of localization transition at nonzero $W$ as well as exhibit a scaling that is consistent with the SPS hypothesis.

\subsection{Structure of Hamiltonian eigenstates} \label{sec:eigenstates}

We finally investigate the transition through the structure of the Hamiltonian eigenfunctions. This direction of research was pioneered by Dean and Bell~\cite{DF9705000055} and has later become one of the central tools to characterize localization transitions in both non-interacting as well as interacting systems.

For a {\it single-particle eigenstate} $|\alpha\rangle$, we calculate its participation entropy and the corresponding average as
\begin{equation} \label{def:Sp}
    S_{\rm P}^{(\alpha)} = \ln \sum\limits_{i=1}^V \left|\psi_{i, \alpha}\right|^4, \hspace{5mm} \overline{S_{\rm P}} = \langle \langle S_{\rm P}^{(\alpha)}\rangle_\alpha \rangle \;,
\end{equation}
where $\psi_{i,\alpha} = \langle i|\alpha\rangle$ is the wavefunction coefficient in the site-occupational basis, $\langle ... \rangle_\alpha$ denotes averaging over single-particle eigenstates, and $\langle ... \rangle$ is the averaging over Hamiltonian realizations.
The protocol of averaging is analogous to the one applied in calculations of the average level spacing ratio $r$ in Sec.~\ref{sec:spectral}.

\begin{figure}[!]
\centering
\includegraphics[width=1.00\columnwidth]{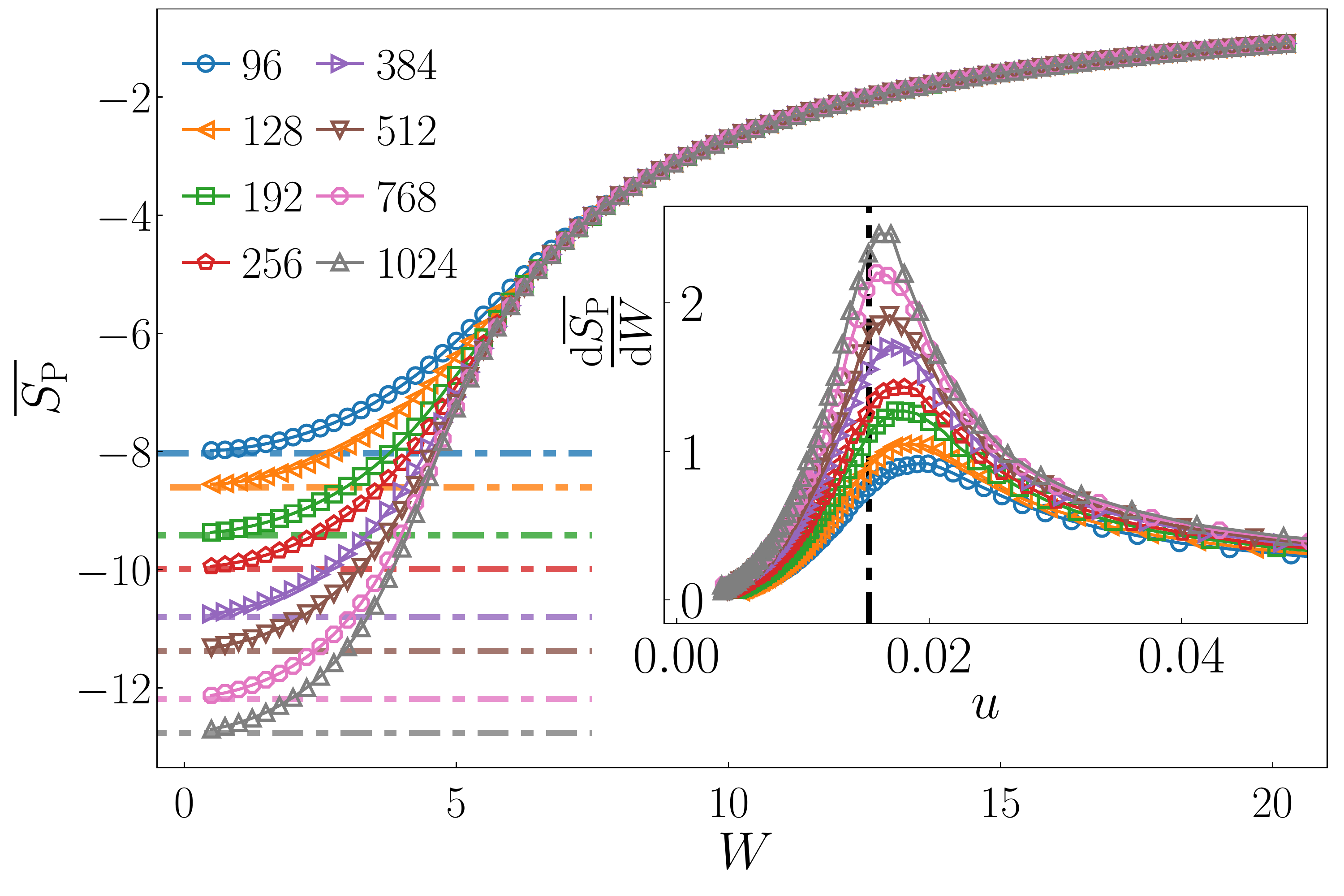}
\caption{
Main panel: average participation entropy $\overline{S_\mathrm{P}}$, see Eq.~(\ref{def:Sp}), versus the disorder $W$. 
The average is performed over 500 single-particle eigenstates centered at the mean energy of the spectrum, and at least 1500 Hamiltonian realizations (or 350 realizations for the largest two system sizes).
Results are shown at different linear system sizes $L=\sqrt{V}$ as indicated in the legend.
Dashed horizontal lines show the corresponding GOE result $\ln (3/V)$ valid in the low-$W$ limit.
Inset: scaling of the derivative
$\frac{{\rm d}\overline{S}_\mathrm{P}}{\mathrm{d}W}$ as a function of the effective parameter $u$.
Dashed vertical line is located at $u^*=0.0153$.
}
\label{fig4}
\end{figure}

In an ideally localized case with the wavefunction equaling unity on a single site while vanishing elsewhere, we have $\overline{S_{\rm P}}=0.$
In the main panel of Fig.~\ref{fig4}, we plot $\overline{S_{\rm P}}$ versus $W$ and we indeed observe an approach towards this limit at large $W$.
In the opposite regime, the wave functions are uniformly spread over the lattice volume and hence the wave function coefficients are of the order $L^{-D/2}$, where $D$ is the lattice dimension.
In particular, if the wavefunction is chaotic and belongs to the GOE universality, one expects $S_{\rm P}^{(\rm GOE)} = \ln (3/V)$~\cite{edwards_thouless_72}.
These values are indicated as horizontal lines in the main panel of Fig.~\ref{fig4}, and we see that the actual numerical results approach these predictions in the limit $W\to 0$.

\begin{figure*}[!]
\centering
\includegraphics[width=2.00\columnwidth]{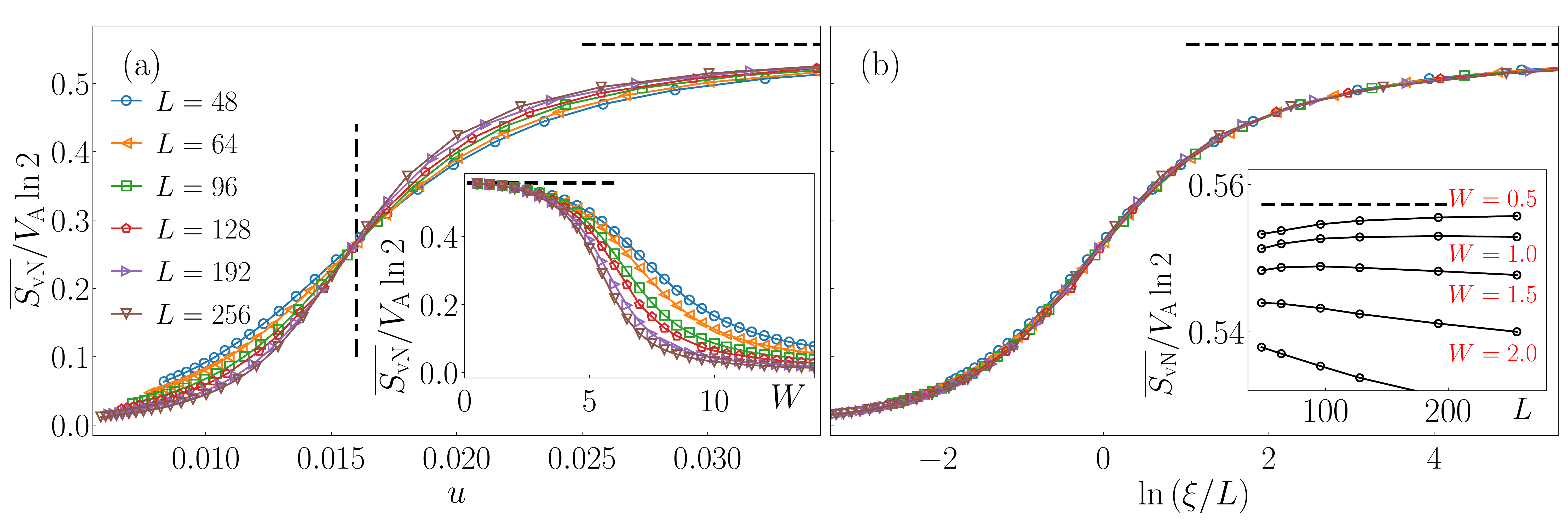}
\caption{
Volume-law coefficient $\overline{S_\mathrm{vN}}/\left(V_A \ln 2\right)$ of the average bipartite entanglement entropy of many-body eigenstates, see Eq.~(\ref{eq:def_SvN}) and the text below.
We partition the 2D lattice in two equal parts, i.e., $V_A = V/2$.
Results are shown for different system sizes $L=\sqrt{V}$, as indicated in the legend.
For $L\leq 128,$ we performed averaging over $N_m=1000$ many-body eigenstates while for the largest two systems we took $N_m=512$ and $N_m=64$ many-body eigenstates, respectively. Additionally, we averaged over approximately $100$ Hamiltonian realizations.
The horizontal dashed lines (in all panels) indicate the maximal value $\overline{{\cal S}_{\rm vN}}/(V_A \ln 2)$ for quantum-chaotic quadratic Hamiltonians, see Eq.~\eqref{eq:def_SvN_ave}.
(a) $\overline{S_\mathrm{vN}}/\left(V_A \ln 2\right)$ as a function of $W$ in the inset, and as a function of the effective parameter $u$ in the main panel.
The scale-invariant point at $u^*=0.016$ is denoted by the vertical dashed line.
(b) Casting $\overline{S_\mathrm{vN}}/\left(V_A \ln 2\right)$ as a function of $\ln(\xi/L)$ yields a remarkable scaling collapse of the data, in analogy with other studied indicators of localization.
The inset shows the size dependence of $\overline{S_\mathrm{vN}}/\left(V_A \ln 2\right)$ at disorder strengths between $W=0.5$ (topmost curve) and $W=2.5$ (lowermost curve) with a stepsize of $\delta W=0.5$.
}
\label{fig5}
\end{figure*}

While our numerical results are in good agreement with both limiting cases discussed above, extracting a quantitative criterion for the localization transition appears less straightforward. As shown in Ref.~\cite{suntajs_prosen_21} for the 3D Anderson model, a very accurate estimate for the critical point $W_c$ can be obtained when studying the behavior of the derivative ${\rm d}\overline{S}_{\rm P} /{\rm d}W$ as a function of $W.$ In the vicinity of the transition, the derivative displays a characteristic peak which becomes both sharper and closer to $W_c$ upon increasing the system size.

In the inset of Fig.~\ref{fig4} we study the derivative ${\rm d}\overline{S}_{\rm P} /{\rm d}W$ in the 2D Anderson model.
We find that a qualitatively similar behavior as in the 3D Anderson model can be obtained once we cast the data as a function of the effective parameter $u$ from Eq.~(\ref{def_u}).
In this case, a sharp peak in ${\rm d}\overline{S}_{\rm P} /{\rm d}W$ emerges in the vicinity of the scale-invariant point $u^*$.
A vertical line in the inset of Fig.~\ref{fig4} denotes the value $u^*=0.0153$ obtained from the analysis of the average level spacing ratio $r$ in Fig.~\ref{fig2}, and it provides a rather accurate prediction for the location of the peak in ${\rm d}\overline{S}_{\rm P} /{\rm d}W$.

We complement the above results by studying properties of {\it many-body eigenstates} $|m\rangle$.
We focus on the entanglement entropies of highly excited eigenstates, which have been recently proposed as a useful measure to distinguish between many-body and single-particle quantum chaos, and the absence thereof~\cite{leblond_mallayya_19, lydzba_rigol_vidmar_prl_2021_PhysRevLett.125.180604, PRXQuantum.3.030201}.

We consider the von Neumann entanglement entropy of the many-body eigenstates $\ket{m}$ of the 2D Anderson model, where $|m\rangle$ are constructed as products of randomly selected single-particle eigenstates.
For a bipartition of a lattice into two connected subsystems $A$ (our system of interest) and $B,$ the von Neumann entanglement entropy of $\ket{m}$ is defined as
\begin{equation}\label{eq:def_SvN}
    S_{\rm vN}^{(m)} = -{\rm Tr}\{\hat{\rho}_A \ln \hat{\rho}_A\}\;,
\end{equation}
where the reduced density matrix of subsystem $A$ is $\hat{\rho}_A = {\rm Tr}_B \{\hat{\rho}_m\}$, and $\hat{\rho}_m = \ket{m}\bra{m}.$
For a subsystem $A$ with volume (the number of lattice sites) $V_A,$ we define the subsystem fraction $f = V_A / V$, with $V = V_A + V_B.$
We then average $S_{\rm vN}^{(m)}$ over different many-body eigenstates $\ket{m}$, as well as over different Hamiltonian realizations, to obtain the average von Neumann eigenstate entanglement entropy $\overline{S_{\rm vN}}$.
Details about the implementation of $\overline{S_{\rm vN}}$ are given in Appendix~\ref{sec:app_SvN}.

For the many-body eigenstates of quantum-chaotic quadratic Hamiltonians, a closed-form analytic expression for the average von Neumann bipartite eigenstate entanglement entropy has recently been introduced~\cite{lydzba_rigol_vidmar_prl_2021_PhysRevLett.125.180604} (see also~\cite{bianchi_hackl_21, PRXQuantum.3.030201}).
The value for bipartitions in two equal parts, i.e., at $f=1/2$, equals
\begin{equation}\label{eq:def_SvN_ave}
    \overline{{\cal S}_{\rm vN}} = \left(2 - 1/\ln 2\right)V_A\ln 2 \approx 0.5573 \, V_A\ln 2 \;.
\end{equation}
Most importantly, the volume-law coefficient $\overline{{\cal S}_{\rm vN}}/(V_A \ln 2)$ in quantum-chaotic quadratic systems at $f=1/2$ is much lower than the maximal value 1 encountered in quantum-chaotic interacting systems~\cite{vidmar_rigol_17, PRXQuantum.3.030201}.
In this context, the eigenstate entanglement entropy is a powerful probe to distinguish between single-particle quantum chaos, which emerges in quantum-chaotic quadratic systems, and many-body quantum chaos, which emerges in most interacting systems.
The validity of Eq.~(\ref{eq:def_SvN_ave}) has been tested numerically in several quantum-chaotic quadratic models, ranging from the quadratic SYK models~\cite{lydzba_rigol_vidmar_prl_2021_PhysRevLett.125.180604, lydzba_rigol_21, liu_chen_18} and the 3D Anderson model at weak disorder~\cite{lydzba_rigol_21} to chaotic tight-binding billiards~\cite{PhysRevE.106.034118}, finding excellent agreement.
Here we use the result in Eq.~(\ref{eq:def_SvN_ave}) as a reference point for the maximal entanglement entropy in the quantum-chaos to localization transition in the 2D Anderson model.

The inset of Fig.~\ref{fig5}(a) shows the dependence of the volume-law coefficient $\overline{S_{\rm vN}}/(V_A \ln 2)$ on the disorder $W$ for different system sizes $V$.
In the limit $W\to 0$ we observe agreement with single-particle quantum chaos, as given by Eq.~(\ref{eq:def_SvN_ave}), while in the regime of large $W$ the onset of localization is manifested as the absence of volume-law scaling, i.e., $\overline{S_{\rm vN}}/(V_A \ln 2) \to 0$.

The results for $\overline{S_{\rm vN}}/(V_A \ln 2)$ versus $W$ in the inset of Fig.~\ref{fig5}(a) resemble the results for $r$ versus $W$ in the inset of Fig.~\ref{fig2}(a).
They both suggest a tendency of localization transition to occur at decreasing values of $W$ upon increasing the system size $V$, however, they do not allow for an unambiguous determination of the fate of the transition point in the thermodynamic limit.
Nevertheless, it is again much more insightful to plot $\overline{S_{\rm vN}}/(V_A \ln 2)$ as a function of the effective parameter $u$, as shown in the main panel of Fig.~\ref{fig5}(a).
The latter results exhibit clear evidence of a scale-invariant point $u^*$, which shares analogies with the results for $r(u)$ in the main panel of Fig.~\ref{fig2}(a), and the results for $\Delta \Sigma^2(u)$ in the inset of Fig.~\ref{fig3}(b).
In Fig.~\ref{fig5}(a), see the vertical line, we obtain the value $u^* = 0.016$, which is very close to the value $u^* = 0.0153$ obtained from $r(u)$ in Fig.~\ref{fig2}(a), and the value $u^*=0.0147$ obtained from $\Delta \Sigma^2(u)$ in Fig.~\ref{fig3}(b).

We also test validity of the SPS hypothesis by plotting $\overline{S_{\rm vN}}/(V_A \ln 2)$ versus $\ln(\xi/L)$ in the main panel of Fig.~\ref{fig5}(b), and get an excellent scaling collapse.
These results suggest that both the absence of localization transition at nonzero $W$, as well as validity of the SPS hypothesis, can readily be identified from the scaling properties of the bipartite entanglement entropies.

Finally, we note a remarkable detail about the scaling properties of the considered quantum chaos indicators.
Even though this work established clear evidence in favor of absence of quantum chaos at any nonzero disorder $W$, this does not imply that, by increasing the system size, the results always exhibit a flow towards localization.
An instructive counterexample is shown in the inset of Fig.~\ref{fig5}(b), where the scaling of the volume-law coefficient $\overline{S_{\rm vN}}/(V_A \ln 2)$ is shown as a function of the linear system size $L$.
At sufficiently weak disorder, i.e., at $W \lesssim 1$, we observe a flow towards quantum chaos prediction $\overline{{\cal S}_{\rm vN}}/(V_A \ln 2)$ from Eq.~(\ref{eq:def_SvN_ave}), see the horizontal dashed line.
If there is no localization transition at nonzero disorder in the 2D Anderson model, this flow is an artefact of finite systems and hence it is expected to exhibit a downturn at much larger system sizes.

\section{Conclusion}

In this work we reconsidered the widely studied 2D Anderson model from the perspective of quantum chaos indicators that are commonly studied in the contemporary literature, such as the level spacing ratio, spectral form factor, variances of observables' matrix elements, participation entropy and eigenstate entanglement entropy.
Many features of those indicators may at a first glance hint at the existence of a stable quantum-chaotic regime at weak disorder.
However, a closer inspection reveals two important observations.

The first observation are remarkable scaling collapses of quantum chaos indicators.
In particular, they exhibit two properties:
emergence of a scale invariant point at a fixed value of an effective parameter $u=1/(W\ln V)$, and scaling collapses consistent with the single-parameter scaling hypothesis.
This confirms validity of the latter at essentially all disorder strengths.

The second observation is that the finite-size analysis of quantum chaos indicators at weak disorder may yield misleading conclusions about the fate of localization transition in the thermodynamic limit.
This was illustrated by a flow of the eigenstate entanglement entropy towards predictions of chaotic systems at weak disorder.
Despite this flow persisted for lattice sizes as large as $10^6$ sites, we interpreted it as a ghost flow that does not survive in the thermodynamic limit.
We referred to the weak disorder regime as prelocalization, which exhibits chaotic properties in finite systems but localization in sufficiently large systems.

We hence conclude that all quantum chaos indicators under investigation are consistent with absence of localization transition at nonzero disorder in the thermodynamic limit, as conjectured in~\cite{abrahams_anderson_79}.
However, at the same time, the results also establish the 2D Anderson model as a toy model to study nontrivial finite-size effects.
In fact, as realized already many years ago~\cite{licciardello_thouless_78, mackinnon1983scaling}, the latter are so severe that samples typically studied in experiments would exhibit a localization transition at nonzero disorder, and hence they cannot be considered to correspond to the thermodynamic limit.
Even more, if one covers the entire surface of planet Earth with a material that is described by the 2D Anderson model~\footnote{Assuming the surface of Earth $S=5 \times 10^{14} m^2$ and the lattice spacing $a=10^{-10} m$, hence $V=5 \times 10^{34}$, which yields $W^*=0.8$ from Eq.~(\ref{def_u_star})}, the localization transition is predicted to occur at $W^* \approx 0.8$.
This calls for considering the statement about the absence of localization transition at nonzero disorder with a grain of salt.

\acknowledgements
We acknowledge discussions with M. Mierzejewski, P. Prelov\v{s}ek, A. Scardicchio and D. Sels.
This work is supported by the Slovenian Research Agency (ARRS), Research core fundings No.~P1-0044 (J.\v S. and L.V.), No.~P1-0402 (T.P.) and No.~J1-1696 (L.V.), and by the European Research Council (ERC) under Advanced Grant 694544 -- OMNES (T.P.). We gratefully acknowledge the High Performance Computing Research Infrastructure Eastern Region (HCP RIVR) consortium \href{https://www.hpc-rivr.si/}{(www.hpc-rivr.si)} and European High Performance Computing Joint Undertaking (EuroHPC JU) \href{https://eurohpc-ju.europa.eu/}{(eurohpc-ju.europa.eu)} for funding this research by providing computing resources of the HPC system Vega at the Institute of Information sciences \href{https://www.izum.si/en/home/}{(www.izum.si)}.


\appendix

\section{Calculation of the spectral form factor} \label{sec:app_sff}

\begin{figure*}[!]
\centering
\includegraphics[width=1.95\columnwidth]{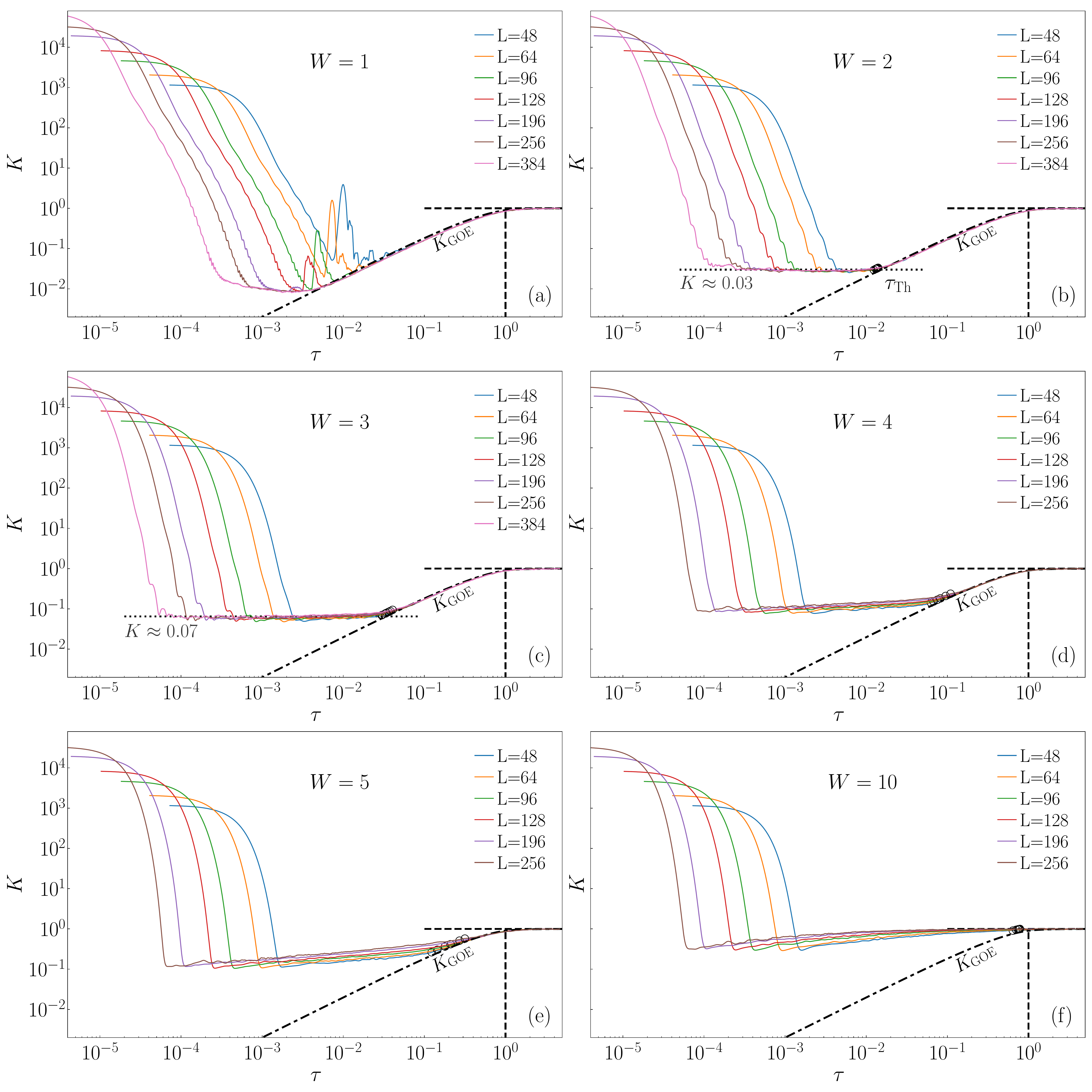}
\caption{
The SFF $K(\tau)$ at six different disorder values $W=1,2,3,4,5,10$ in panels (a)-(f), respectively.
Results are shown for different linear system sizes $L=\sqrt{V}$, as indicated in the legend. 
Dashed-dotted lines denote the GOE result $K_{\rm GOE}(\tau) = 2\tau - \tau\ln(1+2\tau)$ and open circles denote the extracted values of the scaled Thouless time $\tau_{\rm Th}$.
The vertical dashed line is the scaled Heisenberg time $\tau_{\rm H}=1$ and the horizontal dashed line denotes the limiting value $K(\tau) \approx 1$ at large $\tau \gg \tau_{\rm H}$.
The horizontal dotted lines in panels (b) and (c) correspond to $K(\tau)\approx 0.03$ and $K(\tau)\approx 0.07$, respectively.
}
\label{Supp1}
\end{figure*} 

In calculation of the spectral form factor (SFF) $K(\tau)$ from Eq.~(\ref{eq:def_SFF}), and the consequent extraction of the Thouless time, we follow the procedure outlined in Refs.~\cite{suntajs_bonca_20a, suntajs_prosen_21}, which we briefly outline here for convenience. 

We study systems with up to $384\times384$ lattice sites, thus requiring full exact diagonalization of matrices up to $V_0\times V_0,$ where $V_0=384^2 = 147 456.$ 
At a given system size and Hamiltonian realization, we calculate $K(\tau)$ for 5000 times $\tau_i$ in the window $\tau_i \in [1/(2\pi V), 5],$ with $\tau_i$ equidistant in the logarithmic scale.
For system sizes up to $L=128,$ we then average the results over $N_{\rm samples} \approx 1000$ Hamiltonian realizations, while we use $N_{\rm samples} \approx 800, 200, 100$ at $L=196, 256, 384,$ respectively.
Finally, we smoothen out random fluctuations in $K(\tau)$ by calculating a running mean such that each new $K(\tau_i)$ is the average over 50 nearest values of $K(\tau_i),$ and hence the final number of data points is reduced to 4951.

To eliminate the influence of the local density of states, we use the scaled (unfolded) single-particle energy eigenvalues $\{\varepsilon_\alpha\}$ in Eq.~(\ref{eq:def_SFF}), which are obtained by the spectral unfolding.
The main goal of the latter is to transform an ordered set of Hamiltonian single-particle eigenvalues $\{E_\alpha\}$ to an ordered set of unfolded eigenvalues $\{\varepsilon_\alpha\}$ with the local mean level spacing equal to unity at all energy densities. 
The applied protocol of unfolding is identical to the one we used in the 3D Anderson model, see Sec.~3.3.~in Ref.~\cite{suntajs_prosen_21}.

To minimize the effects of spectral edges, we use a filtering function $f(\varepsilon_\alpha)$ in Eq.~\eqref{eq:def_SFF}. As in Refs.~\cite{suntajs_bonca_20a, suntajs_prosen_21}, we filter each unfolded spectrum using a Gaussian filter, $f(\varepsilon_\alpha) = \exp\{\frac{(\varepsilon_\alpha - \bar{\varepsilon})^2}{2(\eta\Gamma)^2}\}.$ Here, $\bar{\varepsilon}$ and $\Gamma^2$ are the average energy and variance, respectively, at a given disorder realization, and $\eta$ a dimensionless parameter controlling the effective fraction of eigenstates included in $K(\tau).$ We used $\eta=0.5$ in our calculations presented in the main text. To ensure proper normalization, yielding $K(\tau \gg 1) \simeq 1$ in general and $K(\tau) \equiv 1$ for Poissonian random spectra, we set the normalization $Z$ in Eq.~\eqref{eq:def_SFF} to $Z=\langle \sum_\alpha |f(\varepsilon_\alpha)|^2\rangle$.

Following Refs.~\cite{suntajs_bonca_20a, suntajs_prosen_21}, we then determine the scaled Thouless time $\tau_{\rm Th}$ by analyzing the deviation of the numerical results from the GOE prediction $K_{\rm GOE}(\tau) = 2\tau - \tau \ln(1+2\tau).$ To that end, we use the deviation measure
\begin{equation}
    \Delta K(\tau) = \left|\log_{10}\frac{K(\tau)}{K_{\rm GOE}(\tau)}\right|; \hspace{5mm} \Delta K(\tau_{\rm Th}) = \epsilon.
\end{equation}
In our calculations, we set $\epsilon=0.08$.
Specifically, since $\Delta K(\tau) \gg \epsilon$ at short times, the Thouless time $\tau_{\rm Th}$ is obtained when $\Delta K(\tau)$ becomes smaller than $\epsilon$, see Fig.~6 of Ref.~\cite{suntajs_bonca_20a}.

In Sec.~\ref{sec:spectral}, we discussed the emergence of a broad plateau in $K(\tau)$ prior to the onset of $\tau_{\rm Th}$, see Fig.~\ref{fig1}(a).
The plateau can be observed at disorder $W=2$ at which finite systems under consideration exhibit seemingly robust signatures of quantum chaos, and the value of the SFF $K(\tau)$ at the plateau is a constant smaller than 1 that appeared to be independent of the system size.
The analysis in Fig.~\ref{fig1}(b) then interpreted these features as finite-size effects.
In Fig.~\ref{Supp1} we show $K(\tau)$ at different disorder values $W=1,2,3,4,5,10$ and different system sizes.
It exhibits several interesting features.
The first is that one can observe signatures of the plateau, or at least a tendency for forming it, at almost all disorders below the crossover regime, $W \lesssim W^*(V) \approx 4$, see Figs.~\ref{Supp1}(a)-\ref{Supp1}(d).
At week disorder $W=1$, see Fig.~\ref{Supp1}(a), one can also observe a peculiar structure similar to beats, which however completely vanish in sufficiently large systems.
When the disorder increases in the regime $W \gtrsim W^*(V)$, the plateau gets replaced by a weak algebraic increase of $K(\tau)$ that is slower that linear, see Figs.~\ref{Supp1}(d) and~\ref{Supp1}(e).
Ultimately, at large disorder the transient regime becomes insignificant, see Fig.~\ref{Supp1}(f), and the SFF $K(\tau)$ complies with the prediction $K(\tau) \approx 1$ expected for systems deep in the localized regime.

\section{Calculation of the von Neumann entanglement entropy} \label{sec:app_SvN}

Here, we briefly outline the main steps performed in our calculations of the bipartite von Neumann entanglement entropy $S_{\rm vN}^{(m)}$ from Eq.~(\ref{eq:def_SvN}) in Sec.~\ref{sec:eigenstates}.
We follow the procedure explained in Ref.~\cite{lydzba_rigol_21}. We denote the single-particle energy eigenkets as $\{\ket{\alpha};\, \alpha = 1, \dots, V\}$, and then construct the many-body eigenkets as $\{\ket{m} = \prod_{\{\alpha_l\}_m}\ket{\alpha_l}; \, m = 1, \dots, 2^V\},$ where $\{\alpha_l\}_m$ represent the $m$-th set of occupied single-particle energy eigenkets.
In practice, we construct a many-body eigenket by randomly selecting each single-particle eigenket with probability 1/2, which means we iterate over all $\ket{\alpha}$ and select or reject them with equal probability.
Hence, while a given many-body eigenket does not necessary belong to the particle-number sector at half filling, the average over many such eigenkets assures that the average particle-number occupation is at, or very close to, half filling.

For particle-number conserving models, such as the Anderson model, all the many-body correlations of an eigenket $\ket{m}$ can be obtained (using Wick's theorem) from the $V\times V$ generalized one-body correlation matrix~\cite{PRXQuantum.3.030201, peschel2003calculation, Peschel_2009}
\begin{equation}\label{eq:def_gen_corr_mat}
    \left(\mathcal{J}_m\right)_{ij} = \bra{m} \hat{c}^\dagger_i\hat{c}_j - \hat{c}_j\hat{c}_i^\dagger \ket{m} = 2(\rho_m)_{ij} - \delta_{ij},
\end{equation}
where $\rho_m$ is the one-body correlation matrix of $\ket{m}$. To calculate the von Neumann entanglement entropy of a many-body eigenket $\ket{m}$, we bipartition the system into connected subsystems $A$ and $B$ and restrict the entries $i,\, j$ of $\mathcal{J}_m$ in Eq.~\eqref{eq:def_gen_corr_mat} to the subsystem $A.$ Then, upon diagonalizing the restricted $\mathcal{J}_m,$ the von Neumann entanglement entropy can be calculated as~\cite{vidmar_hackl_17, PRXQuantum.3.030201}
\begin{equation}
    S_{\rm vN}^{(m)} = -\sum\limits_{i=1}^{V_A}\left(\frac{1+\lambda_i}{2}\ln\left[\frac{1 + \lambda_i}{2} \right] + \frac{1 -\lambda_i}{2}\ln\left[\frac{1 - \lambda_i}{2} \right]\right),
\end{equation}
where $\{\lambda_i\}$ is the spectrum of the restricted $\mathcal{J}_m.$

When calculating averages in the actual numerical calculations, we obtain
\begin{equation}
    \overline{S_{\rm vN}} = \langle \langle S_{\rm vN}^{(m)}\rangle_m \rangle \;,
\end{equation}
where $\langle ... \rangle_m$ denotes the average over randomly chosen many-body eigenkets, and $\langle ... \rangle$ denotes the average over Hamiltonian realizations.

\bibliographystyle{biblev1}
\bibliography{references1,references2,phd_bibliography_2D}

\end{document}